\documentclass[prl,aps,twocolumn,floatfix]{revtex4-1}

\usepackage{graphicx}
\usepackage{epstopdf}

\usepackage{color}

\usepackage[utf8]{inputenc}

\begin{document}
\title{Extended phase diagram of $R$NiC$_2$ family: linear scaling of the Peierls temperature}

\author{Marta Roman, Judyta Strychalska - Nowak, Tomasz Klimczuk, Kamil K. Kolincio} 

\affiliation{Faculty of Applied Physics and Mathematics, Gdansk University of Technology,
Narutowicza 11/12, 80-233 Gdansk, Poland}

\begin{abstract}
Physical properties for the late lanthanide based $R$NiC$_2$ ($R$ = Dy, Ho, Er and Tm) ternary compounds are reported.  All the compounds show antiferromagnetic ground state with the Néel temperature ranging from 3.4 K for HoNiC$_2$ to 8.5 K for ErNiC$_2$. The results of the transport and galvanomagnetic properties confirm a charge density wave state at and above room temperature with transition temperatures $T_{CDW}$ = 284 K, 335 K, 366 K, 394 K for DyNiC$_2$, HoNiC$_2$, ErNiC$_2$ and TmNiC$_2$, respectively. The Peierls temperature $T_{CDW}$ scales linearly with the unit cell volume. A similar linear dependence has been observed for the temperature of the lock-in transition $T_1$ as well. Beyond the intersection point of the trend lines, the lock-in transition is no longer observed. 
In this article we demonstrate an extended phase diagram for $R$NiC$_2$ family.

\end{abstract}

\maketitle

Understanding the interaction between charge density wave (CDW) and other types of ordering such as superconductivity(SC)\cite{Chang2012, daSilva2014, Thampy2017, Caprara2017},  spin density waves (SDW)\cite{Fawcett1988, Jacques2014} and magnetism\cite{Xu2009, Chang2016, Graf2004, Winter2013, Murata2015} is one of the central areas in solid state physics. Recently, a wide interest of the scientists exploring this field has been devoted to two families of ternary compounds: $R_5$Ir$_4$Si$_{10}$ (where $R$ = Y,  Dy,  Ho,  Er,  Tm,  Yb,  or  Lu)\cite{Lalngilneia2015, vanSmaalen2004, Galli2000, Galli2002, Hossain2005, Leroux2013, Singh2004, Sangeetha2012, Kuo2006} and $R$NiC$_2$ (where $R$ = La, Ce, Pr, Nd, Sm, Gd, or Tb)\cite{Kim2013, Prathiba2016, Kim2013, ahmad2015, Shimomura2016, Shimomura2009, Wolfel2010, Hanasaki2011, Hanasaki2017, Kolincio20161, Lei2017, Yamamoto2013, Michor2017}. The uniqueness of those systems originates from the possibility of tuning both the Peierls temperature ($T_{CDW}$) and magnetic ground state by varying the rare-earth element ($R$) \cite{murase2004, Laverock2009, Kolincio2017}. In $R$NiC$_2$ systems, the relevance of the Peierls instability has been confirmed for $R$ = Pr, Nd, Sm, Gd, Tb and Ho, while the LaNiC$_2$ and CeNiC$_2$ compounds do not exhibit any anomalies that could be attributed to CDW. LaNiC$_2$ is found to be a noncentrosymmetric superconductor with critical temperature $T_{SC}$=2.7 K \cite{Wiendlocha2016,lee_superconductivity_1996, Pecharsky1998}. The members of the $R$NiC$_2$ family show a wide range of magnetic orderings originating from the Ruderman-Kittel-Kasuya-Yosida (RKKY) interaction between local magnetic moments and conduction electrons. SmNiC$_2$ orders ferromagnetically with the Curie temperature $T_C$=17.5 K while the rest of $R$NiC$_2$ compounds (with exception of PrNiC$_2$ which shows only a weak magnetic anomaly) exhibit antiferromagnetic transition with the Néel temperature in the range of 2 - 25 K \cite{ONODERA1998_JoMMM}. 
\\
\indent
Although the crystal structure of $R$NiC$_2$ compounds with $R$ belonging to the whole lanthanides series has been determined already, the physical properties of late lanthanides have not been fully studied and the path of the evolution of the charge density wave with $R$ was incomplete. In this paper we extend the phase diagram of $R$NiC$_2$ family to include the late lanthanides ($R$ = Dy, Ho, Er and Tm) with a report of transport, magnetic and galvanomagnetic properties of DyNiC$_2$, HoNiC$_2$, ErNiC$_2$ and TmNiC$_2$ showing Peierls instabilities at and above room temperature. 
\\
\indent
The $R$NiC$_2$ ($R$ = Dy, Ho, Er and Tm) polycrystalline samples were prepared by arc-melting technique followed by annealing at 900$^o$C for 12 days. The detailed procedure was previously described in \cite{Kolincio2017}. Overall loss of weight for DyNiC$_2$ and HoNiC$_2$ after melting and annealing process was negligible ($<$1\%) indicating that the nominal concentration was close to the actual alloying level. For ErNiC$_2$ and TmNiC$_2$ the overall loss was larger ($<$2.5\%) due to high vapor pressure of Er and Tm, therefore appropriate excess of these metals has been added to compensate the deficiency. Phase purity and crystallographic structure of all four samples were confirmed with powder X-ray diffraction (pXRD) measurements (X’Pert PRO-MPD, PANalitycal). All the physical properties  measurements shown in this paper were performed by using commercial Physical Property Measurement System (PPMS, Quantum Design). Electrical resistivity was measured by a standard four-probe method. The Hall effect was measured by reversing the direction of the magnetic field ($\mu_0H$ = 5 T) and the data was antisymmetrized to remove spurious longitudinal magnetoresistance component.
\begin{figure*}[ht!]
\includegraphics[angle=0,width=2.0\columnwidth]{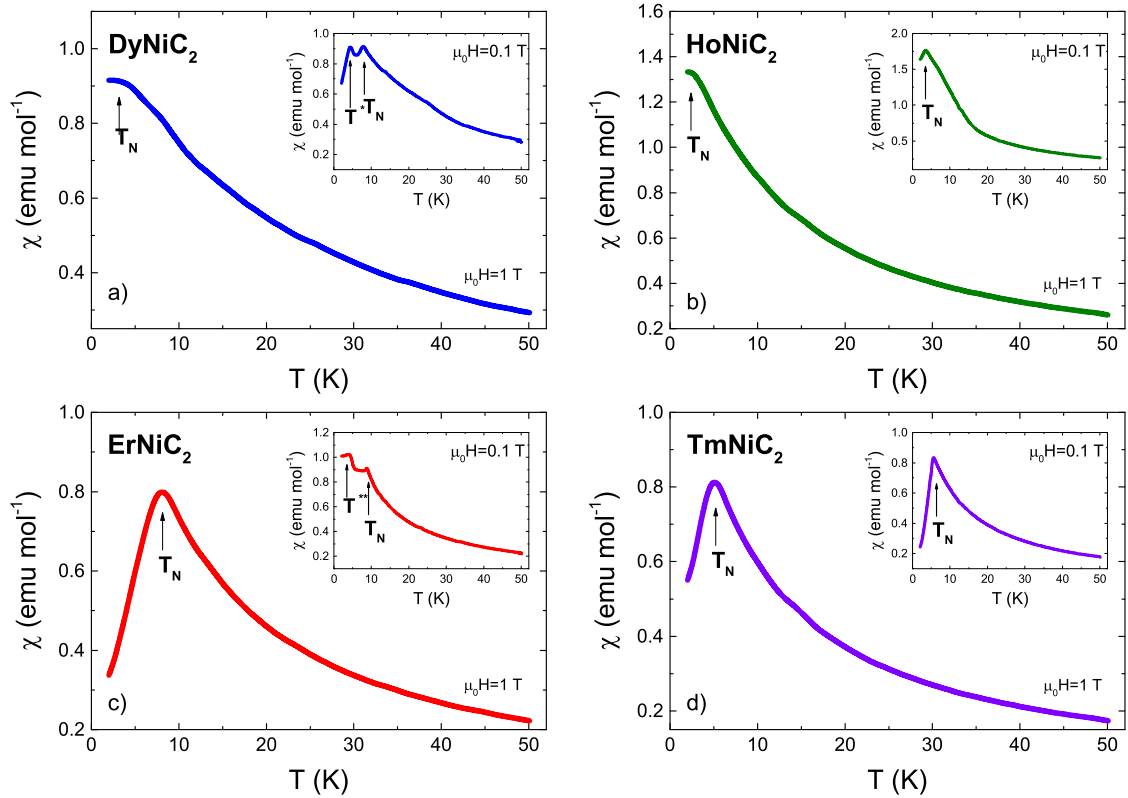}
 \caption{\label{chi} The dc magnetic susceptibility versus temperature $\chi(T)$ of a) DyNiC$_2$, b) HoNiC$_2$, c) ErNiC$_2$ and d) TmNiC$_2$ measured at constant field of 1 T. Insets show $\chi(T)$ measured at 0.1 T magnetic field. Arrows indicate the antiferromagnetic transition at $T_N$ temperature and of the other anomalies $T^*$ and $T^{**}$.}
  \end{figure*}
  
\begin{table}[h!]
 \caption{\label{parameters}Lattice constants, unit cell volume and the figure of merits of the LeBail refinements for DyNiC$_2$, HoNiC$_2$, ErNiC$_2$ and TmNiC$_2$ at room temperature.}
 
 \begin{ruledtabular}\begin{tabular}{ccccc}

   & DyNiC$_2$ &	HoNiC$_2$	& ErNiC$_2$	& TmNiC$_2$ \\
\hline 
$a$ (\AA) & 3.5713(8) &	3.5454(7) &	3.5164(7)	& 3.485(1) \\
$b$ (\AA)	& 4.505(1)	& 4.499(1) &	4.492(1) &	4.486(2) \\
$c$ (\AA)	& 6.038(1)	& 6.026(1)	& 6.014(1) &	5.999(2) \\
$V$ (\AA$^3$)	& 97.151(4)	& 96.109(3) &	94.995(4) & 93.797(5) \\
R$_p$	& 15.0	& 11.4 &	12.0 &	14.9 \\
R$_{wp}$	& 19.4 &	12.9	& 13.9	& 15.0 \\
R$_{exp}$	& 12.66	& 8.04	& 7.35	& 8.87 \\
$\chi^2$	& 2.35	& 2.56	& 3.54	& 2.87 \\

\end{tabular}\end{ruledtabular}
\end{table}

The pXRD measurement revealed that all four samples $R$NiC$_2$ ($R$ = Dy, Ho, Er,  Tm) are single phase and could be indexed in the orthorhombic CeNiC$_2$-type structure with a space group ${Amm2}$. For samples with $R$ = Dy, Ho, Tm small amount of pure unreacted carbon was found. Values of lattice constants (Table \ref{parameters}) were determined from LeBail analysis (see supplementary material \cite{supp}) carried out by using FULLPROF software and are in good agreement with those reported in literature \cite{Kotsanidis1989, Schafer1992}. The decrease of the unit cell volume of $R$NiC$_2$ with $R$ is consistent with the lanthanide contraction.

The dc magnetic susceptibility versus temperature $\chi$(T) of DyNiC$_2$, HoNiC$_2$, ErNiC$_2$ and TmNiC$_2$ is presented in Fig. \ref{chi}.  A sharp drop associated with the antiferromagnetic (AFM) transition  at the Néel temperature ($T_N$ = 3.4 K and 5 K) is clearly observed at 1 T magnetic field for HoNiC$_2$ and TmNiC$_2$ (shown in Fig. \ref{chi} b) and d)). The AFM transition for DyNiC$_2$ and ErNiC$_2$ are more pronounced at 0.1 T magnetic field and the Néel temperature is $T_N$ = 7.8 K and 8.5 K, respectively (shown in the insets of Fig. \ref{chi} a) and c)). The Néel temperature was defined as the maximum of $\chi$(T) and for each compound is in good agreement with previous reports \cite{Onodera1995, Koshikawa1997, Long2001}. For DyNiC$_2$ and ErNiC$_2$ an additional peak at the $\chi$(T) curve at 0.1 T magnetic field is observed at $T^{*}$ = 4 K and $T^{**}$ = 3.6 K (show in the insets of Fig \ref{chi}. a) and c)). These anomalies were previously reported in ref. \cite{Onodera1995, Long2001, Koshikawa1997}. The anomaly seen in ErNiC$_2$ was attributed to another order-order transition and was discussed in \cite{Koshikawa1997}.  

  \begin{figure*}[ht]
  \includegraphics[angle=0,width=2.0\columnwidth]{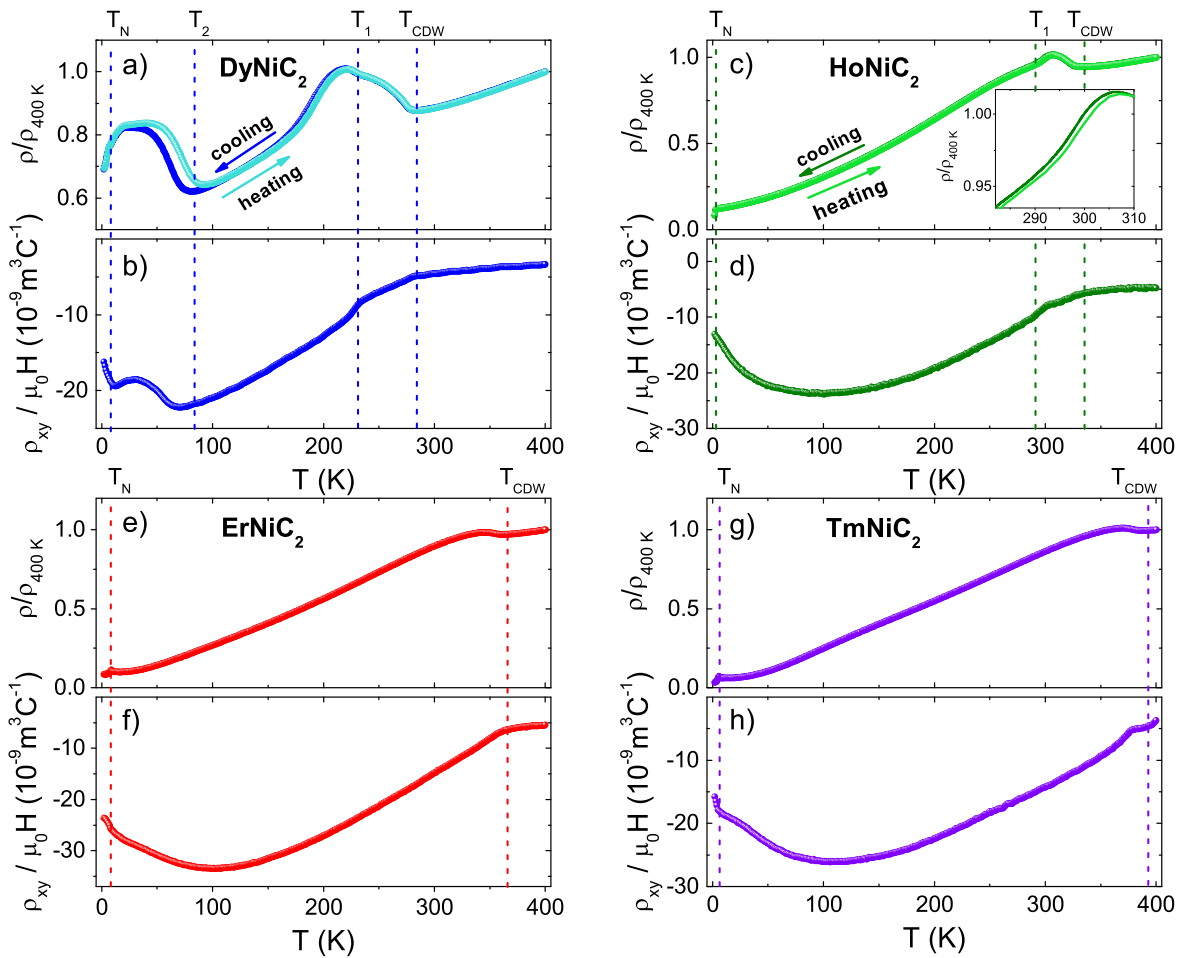}
 \caption{\label{opor} Comparison of the temperature dependence of normalized electrical resistivity and Hall resistivity of DyNiC$_2$ (a and b),  HoNiC$_2$ (c and d), ErNiC$_2$ (e and f) and TmNiC$_2$ (g and h). Arrows indicate the temperature of charge density wave transition $T_{CDW}$ and the Néel temperature $T_N$. Inset shows the expanded view of the hysteresis seen in HoNiC$_2$. Temperature of lock-in transition is marked by $T_1$ and $T_2$ is the temperature of the additional anomaly seen in DyNiC$_2$ (see text for details).}
  \end{figure*}

Fig. \ref{opor} a, c, e and g shows the temperature dependence of the normalized electrical resistivity $\rho/\rho_{400 K}(T)$ for DyNiC$_2$, HoNiC$_2$, ErNiC$_2$ and TmNiC$_2$. At high temperature each compound exhibits a typical metallic character with resistivity lowering as temperature decreases ($d\rho/dT>0$). Upon cooling, an anomaly presenting as a minimum followed by a hump and a crossover to another metallic regime with positive slope of $\rho(T)$ is observed. A similar feature has been reported for other members of the $R$NiC$_2$ family \cite{murase2004} and attributed to a transition into a charge density wave state. The transition temperature was determined from the temperature derivative of resistivity $d\rho/dT$  and denoted $T_{CDW}$ = 284 K, 335 K, 366 K, 394 K for DyNiC$_2$, HoNiC$_2$, ErNiC$_2$ and TmNiC$_2$, respectively.
The transition temperature for HoNiC$_2$ found by us is higher than the value of 317 K reported by Michor \textit{et al.} \cite{Michor2017}. 
In DyNiC$_2$ and HoNiC$_2$, one can notice a small kink at $T_{1}$ = 232 K and 291 K, respectively. This anomaly is accompanied by a small hysteresis, which for HoNiC$_2$ has been shown in an expanded view (inset of Fig. \ref{opor} c). Similar transitions in GdNiC$_2$ and TbNiC$_2$\cite{Shimomura2016} have been identified as lock-in transitions between incommensurate and commensurate CDW states. Next to the analogy with the compounds cited above, another argument suggesting the lock-in character of the transitions seen at $T_1$ in DyNiC$_2$ and HoNiC$_2$ is the existence of a thermal hysteresis - a fingerprint of a first order transition expected by the Ginzburg-Landau approach\cite{McMillan1975}. Temperature resolved  X-ray diffuse scattering experiment is required to unambigously confirm this hypothesis. DyNiC$_2$ shows also an additional feature not present in the other compounds: at $T_2$ = 84 K, one can observe a $\rho(T)$ minimum followed by a hump. This transition also shows a hysteretic behavior, however the hysteresis is significantly wider than the one accompanying the anomaly at $T_1$.  This behavior is not typical for a continuous second order CDW transition expected in the weak coupling scenario with weak lattice distortion. The first order character suggests a significant lattice component of this anomaly as seen in strongly coupled CDW transitions (the key examples are Lu$_5$Ir$_4$Si$_{10}$ and Er$_5$Ir$_4$Si$_{10}$ \cite{Becker1999, Tediosi2009, Jung2003}), or cases in which the Peierls anomaly is triggered by another type of structural distortion as in K$_x$P$_4$W$_8$O$_{32}$\cite{Kolincio20162}. Interestingly, a first order transition can also be observed at a transition between two competing types of ordering as superconductivity and ferromagnetism in ErRh$_4$B$_4$ \cite{Behroozi1983}.
Finally, at $T_N$, established from dc magnetic susceptibility measurements, all the compounds show a sudden decrease of resistivity. This decrease can originate both from the quenching of the spin disorder scattering at the magnetic transition or from partial CDW suppression by antiferromagnetic order as in NdNiC$_2$, GdNiC$_2$ \cite{Yamamoto2013, Kolincio20161, Kolincio2017} or ferromagnetic transition in SmNiC$_2$\cite{hanasaki_magnetic_2012, Lei2017}.

\begin{figure*}[ht!]
  \includegraphics[angle=0,width=1.8\columnwidth]{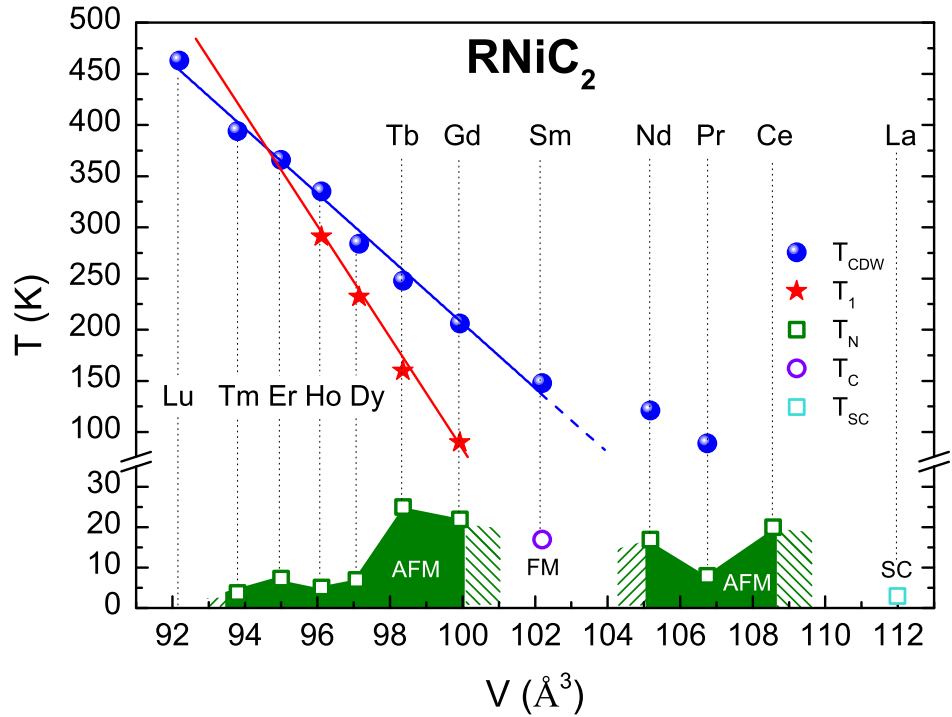}
 \caption{\label{TCDW} Phase diagram for the entire $R$NiC$_2$ series, including late lanthanides ($R$ = Dy, Ho, Er, Tm and Lu). The Peierls ($T_{CDW}$) and the lock-in ($T_1$) transition temperatures are plotted as a function of unit cell volume. The temperatures of the other types of orderings (AFM - $T_N$, FM - $T_C$ and SC - $T_{SC}$)have also been included. The Peierls temperature for LuNiC$_2$ ($T_P$ = 463 K) was revealed by the preliminary resistivity measurements \cite{kolincio2018}.}
  \end{figure*}

The polycrystalline nature of our samples deprives us of the possibility to perform the X-ray diffuse scattering experiment and observe the lattice modulation corresponding to the Peierls transition. Instead, to confirm the CDW character of the observed anomalies, we have studied the galvanomagnetic properties. The  Hall effect is a sensitive probe of the evolution of  the carrier concentration caused by formation of CDW condensate. The Hall resistivity of DyNiC$_2$, HoNiC$_2$, ErNiC$_2$ and TmNiC$_2$ is shown in Fig. \ref{opor} b, d, f and h, respectively. Above $T_{CDW}$, $\rho_{xy}$ is weakly temperature dependent. Below this temperature the Hall resistivity decreases gradually. The downturn of the Hall resistivity at $T_{CDW}$ is a signature of the reduction in carrier concentration and is consistent with the opening of the CDW gap at the Fermi surface. It shall be noted that for TmNiC$_2$ the anomaly in Hall effect is seen at a temperature lower by several K than the minimum in resistivity. For DyNiC$_2$ and HoNiC$_2$ an inflection of $\rho_{xy}$ is visible at $T_1$.  As temperature is decreased further, the Hall resistivity in Ho, Er and Tm bearing compounds reaches a broad minimum and increases as $T$ approaches $T_N$. This trend is continued below the magnetic ordering temperature where a sudden upturn of the $\rho_{xy}$ is observed. In magnetic materials, next to the normal Hall effect ($R_0$) one should also consider the anomalous component of $\rho_{xy}$ (in equation \ref{Hall_eq} represented by $R_S$):
\begin{equation}
\label{Hall_eq}
\rho_{xy}=R_0\mu_0H+4\pi R_SM,
\end{equation} where $M$ is the magnetization.
The $\rho_{xy}$ increase can be then attributed both to the magnetic field induced suppression of CDW concomitant with the release of previously condensed electrons and to the anomalous Hall effect. In a previous study, we have shown that both ingredients of $\rho_{xy}$ are responsible for a similar upturn of Hall resistivity in NdNiC$_2$ and  GdNiC$_2$\cite{Kolincio20161, Kolincio2017}.
For DyNiC$_2$, the $\rho_{xy}(T)$ shows more complex character. In addition to the features discussed above, the Hall resistivity initially decreasing below $T_{CDW}$ reaches a narrow minimum at $T_2$ corresponding to the minimum seen in resistivity. Between $T_2$ and $T_N$ the Hall resistivity reveals a local hump. This behavior confirms the relevance of the electronic component of the transition at $T_2$, coupled with the structural one. The upturn of Hall resistivity can originate from partial destruction of the CDW or alternatively, from nesting of another portion of the Fermi surface and opening hole pockets. Note, that at this temperature no anomaly is observed in magnetic properties. Eventually, at $T_N$, $\rho_{xy}$ increases similarly to the behavior of other studied compounds. The detailed analysis of the anomalies observed for DyNiC$_2$, as well as the detailed analysis of the Hall effect will be continued in a future article.

Fig. \ref{TCDW} depicts the CDW transition temperatures ($T_{CDW}$) for the members of the $R$NiC$_2$ family plotted as a function of unit cell volume. This plot extends the phase diagram previously proposed by Shimomura \textit{et al.} \cite{Shimomura2016}. The authors of the ref. \cite{Shimomura2016} found a linear behavior of the Peierls temperature up to $R$ = Tb. The Pr and Nd bearing compounds were found to deviate slightly from the linear scaling. Here we demonstrate that in agreement with the prediction of Shimomura \textit{et al.}, a linear trend holds for the heavy lanthanides based $R$NiC$_2$ compounds - DyNiC$_2$, HoNiC$_2$ ErNiC$_2$, TmNiC$_2$ studied in this paper and LuNiC$_2$, for which the Peierls temperature of 463 K has been recently revealed by high temperature resistivity measurements \cite{kolincio2018}. Furthermore, we have found that the temperature corresponding to the possible lock-in transition ($T_1$) also scales linearly with the cell volume. Both trend lines intersect near the position of $R$ = Er, where the additional CDW crossover is no longer observed. 
Increase of the Peierls temperature in $R$NiC$_2$ for heavy lanthanides cannot be directly attributed to the increase of the effective low dimensionality as for example in the family of monophosphate tungsten bronzes \cite{Roussel2001, Rotger2004, Schlenker1995}, where the Peierls temperature was significantly enhanced with the separation of conducting layers. In $R$NiC$_2$ family, due to the  lanthanides contraction, the distance between Ni chains (responsible for the charge density wave) \cite{Wolfel2010} decreases with the atomic number of the rare earth metal. Therefore, the mechanism responsible for the enhancement of $T_{CDW}$ could be associated with increase of the interchain coupling or evolution of the band structure, which becomes more favorable for nesting for heavy lanthanides. Interestingly, in contrast to $R$NiC$_2$, in the family of $R_5$Ir$_4$Si$_{10}$, for $R$ ranging from Dy to Lu, $T_{CDW}$ increases with the rare earth ions size\cite{Kuo2003}.

In this article we report the results of powder X-ray diffraction, dc magnetic susceptibility, transport and galvanomagnetic measurements performed on DyNiC$_2$, HoNiC$_2$, ErNiC$_2$ and TmNiC$_2$. The antiferromagnetic transitions ($T_N$ = 7.8 K, 8.5 K, 3.4 K, 5 K for Dy, Ho, Er and Tm, respectively) are in good agreement with previous reports. The charge density wave state for studied compounds is revealed by transport and Hall measurements. The CDW formation temperature is: $T_{CDW}$ = 284 K, 335 K, 366 K, 394 K for DyNiC$_2$, HoNiC$_2$, ErNiC$_2$ and TmNiC$_2$, respectively. These results allowed us to construct the extended and likely completed phase diagram for $R$NiC$_2$ family (including $R$ = Dy, Ho, Er, Tm and Lu). Moreover, we have discovered that $T_{CDW}$ follows a remarkably linear scaling with unit cell volume of the $R$NiC$_2$ for rare earths from Sm to Lu. It was also found that the lock-in transition temperature also obeys a linear dependence.  Beyond the intersection of these trend lines, the lock in transition is no longer observed suggesting the commensurate character of the charge density wave in ErNiC$_2$ and TmNiC$_2$. Diffraction experiments performed with single crystals would be essential to prove this hypothesis. Calculations of the electronic structure are also required to study the enhancement of the Fermi surface nesting for the late lanthanides. It seems to be of particular interest to explore the mechanism behind the linear scaling of $T_{CDW}$.

\begin{acknowledgments}
 Authors gratefully acknowledge the financial support from National Science Centre (Poland), grant number:  UMO-2015/19/B/ST3/03127. We also thank R. Daou and T. Miruszewski for fruitful discussions.
\end{acknowledgments}
%
\end{document}